\documentclass[aps,prd,onecolumn,preprintnumbers,groupedaddress,showpacs,nofootinbib,amssymb]{revtex4}
\usepackage{graphicx}
\usepackage{amsmath}
\usepackage{amssymb}
\usepackage{amsfonts}

\begin{document}

\def\pp{{\, \mid \hskip -1.5mm =}}
\def\cL{{\cal L}}
\def\be{\begin{equation}}
\def\ee{\end{equation}}
\def\bea{\begin{eqnarray}}
\def\eea{\end{eqnarray}}
\def\tr{\mathrm{tr}\, }
\def\nn{\nonumber \\}
\def\e{\mathrm{e}}

\title{Non-singular modified gravity unifying inflation with late-time
acceleration
and universality of viscous ratio bound in $F(R)$ theory}

\author{Shin'ichi Nojiri$^{1,2}$
and Sergei D. Odintsov$^3$\footnote{Also at Tomsk State Pedagogical
University}}

\affiliation{ $^1$ Department of Physics, Nagoya University, Nagoya 464-8602,
Japan \\
$^2$ Kobayashi-Maskawa Institute for the Origin of Particles and
the Universe, Nagoya University, Nagoya 464-8602, Japan \\
$^3$Instituci\`{o} Catalana de Recerca i Estudis Avan\c{c}ats
(ICREA) and Institut de Ciencies de l'Espai (IEEC-CSIC), Campus
UAB, Facultat de Ciencies, Torre C5-Par-2a pl, E-08193 Bellaterra
(Barcelona), Spain}

\begin{abstract}

The review of basic cosmological properties of four-dimensional 
$F(R)$-gravity, including 
FRW equations of motion and its accelerating solutions, generalized fluid 
and scalar-tensor representation of the theory is done.
Cosmological reconstruction equation is written and conditions for 
stability of cosmological solution are discussed.
The overview of realistic $F(R)$-models unifying inflation with dark 
energy epoch is made. The avoidance of finite-time future singularities in 
such theories via the introduction of $R^2$-term is studied.
New realistic non-singular $F(R)$-gravity unifying early-time inflation 
with late-time acceleration is presented. The exit from inflationary era 
in such model may be caused by the gravitational scenario.
It is demonstrated that five-dimensional $F(R)$-gravity considered as 
non-perturbative stringy effective action leads to universal relation for 
viscous bound ratio.

\end{abstract}


\maketitle

\section{Introduction \label{Sec1}}

It is widely accepted that the universe evolution passed at least two
accelerating epochs: the early-time inflation and current late-time
acceleration. The striking similarity between these two accelerating eras
indicates that their origin is caused by the same fundamental scenario.
The most natural and elegant explanation of the known universe evolution in
unified manner may be given within modified gravity approach.
It does not request any extra dark components like inflaton, dark energy and
dark matter which should be implemented into the theory.
The role of these dark components is played by the gravitational theory which
changes its functional form at different curvature scales.
For instance, at the early universe the gravitational theory contains higher
derivatives terms which cause the early-time inflation.
During radiation/matter dominance higher derivatives terms are negligible if
compare with standard General Relativity.
Meanwhile, current dark energy epoch is due to the gravitational terms which
dominate at very low curvature.
The corresponding review on modified gravities unifying the early-time
inflation and late-time acceleration is given in ref.~\cite{Nojiri:2006ri}.
The easiest version of modified gravity which naturally unifies the early-time
inflation and late-time acceleration is
$F(R)$-theory \cite{Nojiri:2006ri,Nojiri:2003ft}.
The number of versions of $F(R)$-gravity is known to be
consistent with astrophysical data and local tests (for review,
see \cite{review}).

In the present paper we study different models of $F(R)$-gravity which may
unify the early-time inflation with late-time acceleration in the consistent
way. Moreover, we present new non-singular models for such
unification.
The paper is organized as follows. In the next section the general properties
of $F(R)$-gravity are reviewed. Spatially-flat FRW equations of motion are
presented. These FRW equations are rewritten in the form of standard FRW
equations for General Relativity with generalized $F(R)$-fluid. Simple
accelerating cosmologies are briefly discussed.
Section \ref{Sec3} is devoted to the review of scalar-tensor representation of
$F(R)$-gravity (so-called Einstein frame). It is shown how to bound the
(non-physical) anti-gravity regime in scalar-tensor representation of the
theory. In section \ref{Sec4} we propose general strategy for cosmological
reconstruction of $F(R)$-gravity, i.e. derivation of its functional form which
contains given cosmological solution. Section \ref{Sec5} is devoted to the 
study
of stability of given cosmological solution.
It gives the cosmological solution conditions which define its
stability/instability.
In section \ref{Sec6} we discuss viable $F(R)$-gravities unifying inflation with dark
energy. The conditions for derivation of such models are given. The review of
the known realistic unified $F(R)$-theories is made.
It is shown that some of these models may contain the finite-time future
singularity. Nevertheless, the addition of $R^2$-term which supports the
early-time inflation cures the future singularities in accord with the original
proposal of refs.~\cite{Abdalla:2004sw,Nojiri:2008fk,bamba}. New non-singular
viable $F(R)$-gravity unifying the inflation with dark energy is proposed and
investigated. It is shown that such theory has the instable inflationary era.
The decay of the inflationary era is caused by the structure of gravitational
theory.
Section \ref{Sec7} is devoted to the study of different, 
five-dimensional application of $F(R)$-gravity. Here, we
discuss five-dimensional $F(R)$-gravity as some non-perturbative stringy
gravity in frames of AdS/CFT correspondence. It is proved that for such theory
like for five-dimensional General Relativity the viscous ratio bound is
universal. Some summary and outlook are given in last section.

\section{General properties of $F(R)$-gravity \label{Sec2}}

Let us briefly review general properties of four-dimensional $F(R)$-gravity
which is known to be realistic candidate for the unification of early-time
inflation with late-time acceleration (for review, see \cite{Nojiri:2006ri}).
The starting action is chosen to be:
\be
\label{JGRG7}
S_{F(R)}= \int d^4x \sqrt{-g} \left(\frac{F(R)}{2\kappa^2} +
\mathcal{L}_\mathrm{matter} \right)\, .
\ee
For general $F(R)$-gravity, one can define an effective equation of state
(EoS) parameter.
The spatially-flat FRW equations in Einstein gravity coupled with perfect fluid
are:
\be
\label{JGRG11}
\rho=\frac{3}{\kappa^2}H^2 \, ,\quad p= - \frac{1}{\kappa^2}\left(3H^2 + 2\dot
H\right)\, .
\ee
For modified gravity, one may define an effective EoS
parameter as follows:
\be
\label{JGRG12}
w_\mathrm{eff}= - 1 - \frac{2\dot H}{3H^2} \, .
\ee

The equation of motion for modified $F(R)$-gravity
is given by
\be
\label{JGRG13}
\frac{1}{2}g_{\mu\nu} F(R) - R_{\mu\nu} F'(R) - g_{\mu\nu} \Box F'(R)
+ \nabla_\mu \nabla_\nu F'(R) = - \frac{\kappa^2}{2}T_{\mathrm{matter}\,
\mu\nu}\, .
\ee
Assuming spatially-flat FRW universe,
\be
\label{JGRG14}
ds^2 = - dt^2 + a(t)^2 \sum_{i=1,2,3} \left(dx^i\right)^2\, ,
\ee
the FRW-like equations are given by
\bea
\label{JGRG15}
&& \rho_\mathrm{eff} = \frac{3}{\kappa^2}H^2 \, ,\quad
p_\mathrm{eff} = - \frac{1}{\kappa^2}\left(3H^2 + 2\dot
H\right)\, , \\
\label{rhoE}
&& \rho_\mathrm{eff} = \frac{1}{\kappa^2}\left(-\frac{1}{2}f(R) + 3\left(H^2 +
\dot H\right) f'(R)
  - 18 \left(4H^2 \dot H + H \ddot H\right)f''(R)\right) +
\rho_\mathrm{matter}\, ,\\
\label{prsr}
&& p_\mathrm{eff} = \frac{1}{\kappa^2}\left(\frac{1}{2}f(R) - \left(3H^2 + \dot
H \right)f'(R)
+ 6 \left(8H^2 \dot H + 4{\dot H}^2
+ 6 H \ddot H + \dddot H \right)f''(R) + 36\left(4H\dot H + \ddot
H\right)^2f'''(R) \right) \nn
&& \qquad + p_\mathrm{matter}\, .
\eea

There may be found several (often exact) solutions of (\ref{JGRG15}).
Nevertheless, due to presence of higher derivatives and non-linear terms, the
structure of exact solutions is much more complicated than in General
Relativity. Without any matter, assuming that the Ricci tensor is covariantly
constant,
that is, $R_{\mu\nu}\propto g_{\mu\nu}$, Eq.~(\ref{JGRG13}) reduces to the
algebraic equation \cite{Cognola:2005de}:
\be
\label{JGRG16}
0 = 2 F(R) - R F'(R)\, .
\ee
If Eq.~(\ref{JGRG16}) has a solution, the Schwarzschild
(or Kerr) - (anti-)de
Sitter or de Sitter space is an exact vacuum solution.

When $F(R)$ behaves as $F(R) \propto R^m$ and there is no matter, there appears
the following solution:
\be
\label{JGRG17}
H \sim \frac{-\frac{(m-1)(2m-1)}{m-2}}{t}\, ,
\ee
which gives the following effective EoS parameter:
\be
\label{JGRG18}
w_\mathrm{eff}=-\frac{6m^2 - 7m - 1}{3(m-1)(2m -1)}\, .
\ee

When $F(R) \propto R^m$ again but if the matter with a constant EoS
parameter $w$ is included,
one may get the following solution:
\be
\label{JGRG19}
H \sim \frac{\frac{2m}{3(w+1)}}{t}\, ,
\ee
and the effective EoS parameter is given by
\be
\label{JGRG20}
w_\mathrm{eff}= -1 + \frac{w+1}{m}\, .
\ee
This shows that modified gravity may describe early/late-time universe
acceleration which could reproduce the quintessence-like or phantom-like
behaviour. Realistic non-linear $F(R)$ models are discussed below.

\section{Scalar-tensor description of $F(R)$-gravity \label{Sec3}}

It is well-known that one can rewrite $F(R)$-gravity action in the
scalar-tensor form.
Introducing the auxiliary field $A$, we rewrite the action (\ref{JGRG7}) of
the $F(R)$-gravity as:
\be
\label{JGRG21}
S=\frac{1}{2\kappa^2}\int d^4 x \sqrt{-g} \left\{F'(A)\left(R-A\right) +
F(A)\right\}\, .
\ee
By the variation over $A$, one obtains $A=R$. Substituting $A=R$ into
the action (\ref{JGRG21}),
one can reproduce the action (\ref{JGRG7}). Furthermore, we rescale the
metric in the following way (conformal transformation):
\be
\label{JGRG22}
g_{\mu\nu}\to \e^\sigma g_{\mu\nu}\, ,\quad \sigma = -\ln F'(A)\, .
\ee
Hence, the Einstein frame action is obtained:
\bea
\label{JGRG23}
S_E &=& \frac{1}{2\kappa^2}\int d^4 x \sqrt{-g} \left( R -
\frac{3}{2}g^{\rho\sigma}
\partial_\rho \sigma \partial_\sigma \sigma - V(\sigma)\right) \, ,\nn
V(\sigma) &=& \e^\sigma g\left(\e^{-\sigma}\right)
   - \e^{2\sigma} f\left(g\left(\e^{-\sigma}\right)\right) = \frac{A}{F'(A)} -
\frac{F(A)}{F'(A)^2}\, .
\eea
Here $g\left(\e^{-\sigma}\right)$ is given by solving the equation
$\sigma = -\ln\left( 1 + f'(A)\right)=- \ln F'(A)$ as
$A=g\left(\e^{-\sigma}\right)$.
Due to the scale transformation (\ref{JGRG22}), there appears a coupling
of the scalar field $\sigma$ with usual matter when matter is included.
The mass of $\sigma$ is given by
\be
\label{JGRG24}
m_\sigma^2 \equiv \frac{3}{2}\frac{d^2 V(\sigma)}{d\sigma^2}
=\frac{3}{2}\left\{\frac{A}{F'(A)} - \frac{4F(A)}{\left(F'(A)\right)^2} +
\frac{1}{F''(A)}\right\}\, .
\ee
Unless $m_\sigma$ is very large, there appears the large
correction to the Newton law.
Naively, one expects the order of the mass $m_\sigma$ could be that of the
Hubble rate, that is,
$m_\sigma \sim H \sim 10^{-33}\,\mathrm{eV}$, which is very light and the
correction could be
very large.

In ref.~\cite{Hu:2007nk} (see also ref.~\cite{bat}), a ``realistic'' $F(R)$
model has been proposed. It may be shown
however, that the model has a instability where the large curvature can be
easily produced.
This is the manifestation of the finite-time future singularity.
In the model \cite{Hu:2007nk}, a parameter $m\sim 10^{-33}\, \mathrm{eV}$
with mass dimension is included. The parameter $m$ plays a role of
the effective cosmological constant.
When the curvature $R$ is large enough compared with
$m^2$, $R\gg m^2$, $F(R)$ \cite{Hu:2007nk}
has the following approximate form:
\be
\label{HS1}
F(R) = R - c_1 m^2 + \frac{c_2 m^{2n+2}}{R^n} +
\mathcal{O}\left(R^{-2n}\right)\, .
\ee
Here $c_1$, $c_2$, and $n$ are positive dimensionless constants. Then the
potential $V(\sigma)$
  (\ref{JGRG23}) has the following asymptotic form:
\be
\label{HS2}
V(\sigma) \sim \frac{c_1 m^2}{A^2}\, .
\ee
Hence, the infinite curvature $R=A\to \infty$ corresponds to small value of the
potential
and therefore the large curvature can be easily produced.

Let us assume that when $R$ is large, $F(R)$ behaves as
\be
\label{HS3}
F(R) \sim F_0 R^{\epsilon}\, .
\ee
Here $F_0$ and $\epsilon$ are positive constant. One also assumes $\epsilon>1$
so that this term
dominates if compared with General Relativity. Then the potential $V(\sigma)$
(\ref{JGRG23}) behaves as
\be
\label{HS4}
V(\sigma) \sim \frac{\epsilon -1}{\epsilon^2 F_0 R^{\epsilon -2}}\, .
\ee
Therefore if $1<\epsilon<2$, the potential $V(\sigma)$ diverges when $R\to
\infty$ and therefore
the large curvature is not realized so easily.
When $\epsilon=2$, $V(\sigma)$ takes finite value $1/F_0$ when $R\to \infty$.
As long as $1/F_0$ is large
enough, the large curvature could be prevented.


Note that there appears the anti-gravity regime when $F'(R)$ is negative,
which follows
from Eq.~(\ref{JGRG21}) of ref.~\cite{Nojiri:2003ft} where first $FR)$ model
unifying inflation with dark energy was proposed.
Hence, one should require
\be
\label{FR1}
F'(R) > 0 \, .
\ee
As long as the condition (\ref{FR1}) is satisfied there occurs usual gravity
regime although there might
appear extra force when $m_\sigma^2$ (\ref{JGRG24}) is small.
Note also that
\be
\label{FRV1}
\frac{dV(\sigma)}{dA} = \frac{F''(A)}{F'(A)^3}\left( - AF'(A) + 2F(A) \right)\,
.
\ee
Therefore if
\be
\label{FRV2}
0 = - AF'(A) + 2F(A) \, ,
\ee
the scalar field $\sigma$ is on the local maximum or local minimum of the
potential and
therefore $\sigma$ can be a constant. The condition (\ref{FRV2}) is nothing but
the
condition (\ref{JGRG16}) for the existence of the de Sitter
solution \cite{Cognola:2005de}.
When the condition (\ref{FRV2}) is satisfied, the mass (\ref{JGRG24}) can be
rewritten as
\be
\label{FRV3}
m_\sigma^2 = \frac{3}{2 F'(A)} \left( - A + \frac{F'(A)}{F''(A)} \right)\, .
\ee
Then when the condition (\ref{FR1}) for the exclusion of the anti-gravity
is
satisfied, the mass squared $m_\sigma^2$ is positive and therefore the scalar
field is
on the local minimum if
\be
\label{FRV4}
   - A + \frac{F'(A)}{F''(A)} > 0\, .
\ee
On the other hand, if
\be
\label{FRV5}
   - A + \frac{F'(A)}{F''(A)} < 0\, ,
\ee
the scalar field is on the local maximum of the potential and
the mass squared $m_\sigma^2$ is negative.
As we will see later in (\ref{FR7}), the condition (\ref{FRV4}) is nothing but
the condition
for the stability of the de Sitter space.
Thus, scalar-tensor representation (Einstein frame) for $F(R)$-gravity is
constructed.
Despite the mathematical equivalence of Jordan and Einstein frames pictures
they lead to the theories which are not physically equivalent (for
corresponding discussion, see \cite{Capozziello:2006dj}).
(Or, more exactly, if physics in one frame is matched with observations and
describe the observable accelerating universe then another frame physics
becomes extremely unconventional. For instance, instead of acceleration the
observerver sees the decceleration, or matter couples with dilaton, etc.)
Note also that it is evident how to transform $F(R)$-gravity to the effective
General Relativity with generalized $F(R)$ fluid \cite{Nojiri:2008ku}.
Of course, the generalized fluid is gravitational fluid and it contains
higher derivatives of curvature invariants. This fact is often ignored in the
study of cosmological perturbations in $F(R)$-gravity, which leads to basically
erroneous conclusions or not well-justified approximations where fourth-order
differential equations become second order ones similar to standard General
Relativity cosmological perturbations (see, for instance,
ref.~\cite{DeFelice:2010aj}). For correct treatment the covariant,
higher-derivative
cosmological perturbations theory of $F(R)$-gravity was developed in
ref.~\cite{dunsby1}.
However, it turns out that corresponding higher-derivative equations are much
more complicated than the corresponding ones in General Relativity.

\section{Cosmological reconstruction of modified $F(R)$-gravity \label{Sec4}}

Let us demonstrate that any FRW cosmology may be realized in specific
$F(R)$-gravity. In other words, we propose the general solution for inverse
problem, i.e. the cosmological reconstruction of $F(R)$-gravity (for the
introduction and review, see \cite{rec}).
   Let us rewrite the FRW equation (\ref{JGRG15}) by using a new variable
(which is often called e-folding) instead of the cosmological time $t$,
$N=\ln \frac{a}{a_0}$.
The variable $N$ is related with the redshift $z$ by
\be
\label{Nz}
\e^{-N}=\frac{a_0}{a} = 1 + z\, .
\ee
Since $\frac{d}{dt} = H \frac{d}{dN}$ and therefore
$\frac{d^2}{dt^2} = H^2 \frac{d^2}{dN^2} + H \frac{dH}{dN} \frac{d}{dN}$,
one can rewrite (\ref{JGRG15}) as (for details, see \cite{Nojiri:2009kx})
\be
\label{RZ4}
0 = - \frac{F(R)}{2} + 3 \left( H^2 + H H'\right) F'(R)
   - 18 \left(4 H^3 H' + H^2 \left(H'\right)^2 + H^3 H''\right) F''(R)
+ \kappa^2 \rho_\mathrm{matter}\, .
\ee
Here $H'\equiv dH/dN$ and $H''\equiv d^2 H/dN^2$.
If the matter energy density $\rho_\mathrm{matter}$ is given by a sum of the
fluid
densities with constant EoS
parameter $w_i$, one finds
\be
\label{RZ6}
\rho_\mathrm{matter} = \sum_i \rho_{i0} a^{-3(1+w_i)}
= \sum_i \rho_{i0} a_0^{-3(1+w_i)} \e^{-3(1+w_i)N}\, .
\ee
Let the Hubble rate is given in terms of $N$ via the function $g(N)$ as
\be
\label{RZ7}
H=g(N) = g \left(- \ln\left(1+z\right)\right)\, .
\ee
Then scalar curvature takes the form: $R = 6 g'(N) g(N) + 12 g(N)^2$,
which could be solved with respect to $N$ as
$N=N(R)$. Using (\ref{RZ6}) and (\ref{RZ7}), one can rewrite
(\ref{RZ4}) as
\bea
\label{RZ9}
0 &=& -18 \left(4g\left(N\left(R\right)\right)^3 g'\left(N\left(R\right)\right)
+ g\left(N\left(R\right)\right)^2 g'\left(N\left(R\right)\right)^2
+ g\left(N\left(R\right)\right)^3g''\left(N\left(R\right)\right)\right)
\frac{d^2 F(R)}{dR^2} \nn
&& + 3 \left( g\left(N\left(R\right)\right)^2
+ g'\left(N\left(R\right)\right) g\left(N\left(R\right)\right)\right)
\frac{dF(R)}{dR}
   - \frac{F(R)}{2}
+ \kappa^2 \sum_i \rho_{i0} a_0^{-3(1+w_i)} \e^{-3(1+w_i)N(R)}\, ,
\eea
which constitutes a differential equation for $F(R)$, where the variable
is scalar curvature $R$.
Instead of $g$, if we use $G(N) \equiv g\left(N\right)^2 = H^2$,
the expression (\ref{RZ9}) could be a little bit simplified:
\bea
\label{RZ11}
0 &=& -9 G\left(N\left(R\right)\right)\left(4 G'\left(N\left(R\right)\right)
+ G''\left(N\left(R\right)\right)\right) \frac{d^2 F(R)}{dR^2}
+ \left( 3 G\left(N\left(R\right)\right)
+ \frac{3}{2} G'\left(N\left(R\right)\right) \right) \frac{dF(R)}{dR} \nn
&& - \frac{F(R)}{2}
+ \kappa^2 \sum_i \rho_{i0} a_0^{-3(1+w_i)} \e^{-3(1+w_i)N(R)}\, .
\eea
Note that the scalar curvature is given by $R= 3 G'(N) + 12 G(N)$.
Hence, when we find $F(R)$ satisfying the differential equation
(\ref{RZ9}) or (\ref{RZ11}),
such $F(R)$ theory
admits the solution (\ref{RZ7}). The reconstructed $F(R)$-gravity realizes
any given above cosmological solution. For instance, one can present the forms
of $F(R)$-theory which describe inflation or dark energy or even unify the
early-time inflation with late-time acceleration era. The number of
corresponding examples may be explicitly realized. However, as a rule the
corresponding theory is some complicated non-analytical function expressed in
terms of special functions. Hence, the cosmological reconstruction turns out to
be approximate one. The exact reconstruction of $\Lambda$CDM universe from pure
$F(R)$-gravity often leads to (almost) General Relativity \cite{dunsby}.

\section{Stability of the cosmological solution \label{Sec5}}

In the previous section, we presented a formulation to obtain the form of
$F(R)$ to
reproduce the known solution $H=H(N)$.
Note, however, the solution $H=H(N)$ is not always stable.
In this section, the condition of the stability of such solution is studied.
The FRW equation (\ref{JGRG15}) can be rewritten by using $G=H^2$ and the
e-folding $N$,
as in (\ref{RZ11}):
\bea
\label{FR_2}
0 &=& -9 G(N) \left(4 G'(N) + G''(N) \right) \left. \frac{d^2 F(R)}{dR^2}
\right|_{R=3 G'(N) + 12 G(N)}
+ \left( 3 G(N) + \frac{3}{2} G'(N) \right) \left. \frac{dF(R)}{dR}
\right|_{R=3 G'(N) + 12 G(N)} \nn
&& - \left. \frac{F(R)}{2} \right|_{R=3 G'(N) + 12 G(N)}
+ \kappa^2 \rho_\mathrm{matter} \left(N\right)\, .
\eea
Let us assume the $N$-dependence of $\rho_\mathrm{matter}$ is known as in
(\ref{RZ6}).
Let a solution of (\ref{FR_2}) be $G=G_0(N)$ and consider the perturbation
from the solution:
\be
\label{FR_3}
G(N) = G_0(N) + \delta G(N)\, .
\ee
Then Eq.~(\ref{FR_2}) gives
\bea
\label{FR_4}
0 &=& G_0(N) \left. \frac{d^2 F(R)}{dR^2} \right|_{R=3 G_0'(N) + 12 G_0(N)}
\delta G'' (N)
+ \left\{ 3 G_0(N) \left( 4G_0'(N) + G_0''(N) \right) \left. \frac{d^3
F(R)}{dR^3} \right|_{R=3 G_0'(N) + 12 G_0(N)} \right. \nn
&& \left. + \left( 3 G_0(N) - \frac{1}{2} G_0'(N) \right) \left. \frac{d^2
F(R)}{dR^2} \right|_{R=3 G_0'(N) + 12 G_0(N)}
\right\} \delta G'(N) \nn
&& + \left\{ 12 G_0 (N) \left( 4G_0'(N) + G_0''(N) \right) \left. \frac{d^3
F(R)}{dR^3} \right|_{R=3 G_0'(N) + 12 G_0(N)}
\right. \nn
&& \left. + \left( - 4G_0(N) + 2 G_0'(N) + G_0''(N) \right) \left. \frac{d^2
F(R)}{dR^2} \right|_{R=3 G_0'(N) + 12 G_0(N)}
+ \frac{1}{3} \left. \frac{d F(R)}{dR} \right|_{R=3 G_0'(N) + 12 G_0(N)}
\right\} \delta G \, .
\eea
Note that in this formulation, as it is assumed that the $N$-dependence of
$\rho_\mathrm{matter}$ is known,
one need not consider the fluctuation of $\rho_\mathrm{matter}$. If the
cosmological time $t$ is used
instead of $N$, we need to include the fluctuation of $\rho_\mathrm{matter}$.
Thus, the stability conditions are given by
\bea
\label{FR_5}
&& 6 \left( 4 G_0'(N) + G_0''(N)\right) \left. \frac{d^3 F(R)}{dR^3}
\right|_{R=3 G_0'(N) + 12 G_0(N)}
\left( \left. \frac{d^2 F(R)}{dR^2} \right|_{R=3 G_0'(N) + 12 G_0(N)}
\right)^{-1} + 6
   - \frac{G_0'(N)}{G_0(N)} >0 \, , \\
\label{FR_6}
&& 36 \left( 4G_0'(N) + G_0''(N) \right) \left. \frac{d^3 F(R)}{dR^3}
\right|_{R=3 G_0'(N) + 12 G_0(N)}
\left( \left. \frac{d^2 F(R)}{dR^2} \right|_{R=3 G_0'(N) + 12 G_0(N)}
\right)^{-1} - 12
+ \frac{6 G_0'(N)}{G_0(N)} \nn
&& + \frac{3 G_0''(N)}{G_0(N)} + \frac{1}{G_0(N)} \left. \frac{d F(R)}{dR}
\right|_{R=3 G_0'(N) + 12 G_0(N)}
\left( \left. \frac{d^2 F(R)}{dR^2} \right|_{R=3 G_0'(N) + 12 G_0(N)}
\right)^{-1} >0 \, .
\eea
In case of de Sitter space, where $H$ and therefore $G_0$ and $R=R_0 \equiv 12
G_0$ are constants,
Eq.~(\ref{FR_5}) becomes $6>0$ and trivially
satisfied and Eq.~(\ref{FR_6}) becomes
\be
\label{FR7}
- 12 G_0 + \left. \frac{d F(R)}{dR} \right|_{R=12 G_0}
\left( \left. \frac{d^2 F(R)}{dR^2} \right|_{R=12 G_0} \right)^{-1}
= - R_0 + \left. \frac{d F(R)}{dR} \right|_{R=R_0}
\left( \left. \frac{d^2 F(R)}{dR^2} \right|_{R=R_0} \right)^{-1} >0 \, ,
\ee
which is the standard result.

One may consider the case that a form of $F(R)$ admits two de Sitter solutions.
If one is stable but another
is unstable, there could be a solution that matches the unstable de Sitter
solution to the stable solution.
If the Hubble rate $H$ of the unstable solution is much larger than the Hubble
rate of the stable one,
the unstable solution may correspond to the inflation and the stable one to the
late-time acceleration.
Alternatively, we may directly construct $F(R)$-gravity which admits the
transition from asymptotic de Sitter
universe with the large Hubble rate to another asymptotic de Sitter
universe with the small Hubble rate. Then if the corresponding solution of
$F(R)$-theory satisfies the conditions (\ref{FR_5})
and (\ref{FR_6}), such a transition surely occurs. One should bear in mind
that other scenarios for exit from inflationary phase (for instance, due to
quantum effects) may be proposed as well.

Let us now consider the stability of de Sitter universe satisfying the
condition (\ref{JGRG16}), or equivalently
(\ref{FRV2}). Eq.~(\ref{JGRG16}) can be rewritten as
\be
\label{dS1}
0 = \frac{d}{dR}\left( \frac{F(R)}{R^2} \right) \, .
\ee
Let $R=R_0$ is a solution of (\ref{dS1}). Then $F(R)$ has the following form:
\be
\label{dS2}
\frac{F(R)}{R^2} = f_0 + \tilde f (R) \left( R - R_0 \right)^n \, .
\ee
Here $f_0$ is a constant, which should be positive if $F(R)>0$
and $n$ is an integer greater or equal to two:$n\geq 2$.
We assume the function $\tilde f(R)$ does not vanish at $R=R_0$, $\tilde
f(R_0)\neq 0$.
When $n=2$, it follows
\be
\label{dS3}
- R_0 + \frac{F'(R_0)}{F''(R_0)} = - \frac{\tilde f(R_0)A_0}{f_0 + f(A_0)}\, .
\ee
Eq.~(\ref{FR7}) shows that the de Sitter solution is stable if
\be
\label{dS4}
- f_0 < f(A_0) < 0\, .
\ee
When $n\geq 3$, one gets
\be
\label{dS5}
- R_0 + \frac{F'(R_0)}{F''(R_0)} = 0\, .
\ee
Hence, we need more detailed investigation to check the stability.
One now uses the expression of $m_\sigma^2$ in (\ref{JGRG24}) and
investigates the sign of $m_\sigma$ in the region $R\sim R_0$.
The expression (\ref{FRV3}) cannot be used since
it is only valid in the point $R=R_0$.
Therefore
\be
\label{dS6}
m_\sigma^2 \sim - \frac{3n(n-1) R_0^2\tilde f(R_0)}{2f_0^2}
\left( R - R _0\right)^{n-2} \, .
\ee
Eq.~(\ref{dS6}) shows that when $n$ is an even integer, the de Sitter solution
is stable if $\tilde f(R_0)<0$ but unstable if $\tilde f(R_0)>0$.
On the other hand, when $n$ is an odd integer, the de Sitter solution is
always unstable. Note, however, that when $\tilde f(R_0)<0$ $\left( \tilde
f(R_0)>0 \right)$,
we find $m_\sigma^2 >0$ $\left(m_\sigma^2 < 0\right)$ if $R>R_0$ but
$m_\sigma^2 <0$ $\left(m_\sigma^2 > 0\right)$ if $R<R_0$.
Therefore when $\tilde f(R_0)<0$, $R$ becomes smaller but when
$\tilde f(R_0)>0$, $R$ becomes larger.

Thus, it is developed the scheme which permits to check when the unification of
inflation with dark energy is realistic one because inflarionary stage is
instable.

\section{Viable $F(R)$-gravities unifying inflation with dark energy \label{Sec6}}


\subsection{The known realistic $F(R)$-models unifying inflation with dark
energy}

In refs.~\cite{Nojiri:2007as}, \cite{Nojiri:2007cq}, and \cite{Cognola:2007zu},
viable models of $F(R)$-gravity
   unifying the late-time acceleration and the inflation were proposed.
To construct these models, we have required several conditions:
\begin{enumerate}
\item In order to generate the inflation, one may require
\be
\label{Uf1}
\lim_{R\to\infty} f (R) = - \Lambda_i\, .
\ee
Here $\Lambda_i$ is an effective cosmological constant at the early universe
and therefore
it is natural to assume $\Lambda_i \gg \left(10^{-33}\mathrm{eV}\right)^2$.
For instance, it is natural to have $\Lambda_i\sim 10^{20 \sim 38}\left(
\mathrm{eV}\right)^2$.
\item In order that the current cosmic acceleration could be generated,
the current $f(R)$-gravity is considered to be a small constant, that is,
\be
\label{Uf3}
f(R_0)= - 2\tilde R_0\, ,\quad f'(R_0)\sim 0\, .
\ee
Here $R_0$ is the current curvature $R_0\sim
\left(10^{-33}\mathrm{eV}\right)^2$.
Note that $R_0> \tilde R_0$ due to the contribution from matter. In fact, if
we can regard $f(R_0)$ as an effective cosmological constant, the effective
Einstein equation gives $R_0=\tilde R_0 - \kappa^2 T_\mathrm{matter}$.
Here $T_\mathrm{matter}$ is the trace of the matter energy-momentum tensor.
Note that $f'(R_0)$ need not to vanish exactly. Since the time scale of one-ten
billion years is considered, we only require
$\left| f'(R_0) \right| \ll \left(10^{-33}\,\mathrm{eV}\right)^4$.

Instead of the model corresponding to (\ref{Uf1}), one may consider a model
which satisfies
\be
\label{UU2}
\lim_{R\to\infty} f (R) = \alpha R^m \, ,
\ee
with a positive integer $m>1$ and a constant $\alpha$.
In order to avoid the anti-gravity $f'(R)>-1$, we find $\alpha>0$ and
therefore $f(R)$ should be positive at the early universe.
On the other hand, Eq.~(\ref{Uf3}) shows that $f(R)$ is negative at
the present universe. Therefore $f(R)$ should cross zero in the past.
\item The last condition is
\be
\label{Uf4}
\lim_{R\to 0} f(R) = 0\, ,
\ee
which means that there is a flat space-time solution.
\end{enumerate}


One typical model proposed in ref.~\cite{Nojiri:2007as} and satisfying the
above
conditions is
\be
\label{UU2d}
f(R)= \frac{\alpha R^{2n} - \beta R^n}{1 + \gamma R^n}\, .
\ee
Here $\alpha$, $\beta$, and $\gamma$ are positive constants and $n$ is a
positive integer.
Eq.~(\ref{UU2d}) gives \cite{Nojiri:2007as}
\be
\label{UUU7}
R_0=\left\{ \left(\frac{1}{\gamma}\right)
\left(1+ \sqrt{ 1 + \frac{\beta\gamma}{\alpha} }\right)\right\}^{1/n}\, ,
\ee
and therefore
\be
\label{UU6}
f(R_0) \sim -2 \tilde R_0 = \frac{\alpha}{\gamma^2}
\left( 1 + \frac{\left(1 - \frac{\beta\gamma}{\alpha} \right)
\sqrt{ 1 + \frac{\beta\gamma}{\alpha}}}{2 + \sqrt{ 1 +
\frac{\beta\gamma}{\alpha}}} \right) \, .
\ee
Then it follows
\be
\label{UU9}
\alpha \sim 2 \tilde R_0 R_0^{-2n}\, ,\quad \beta \sim 4 {\tilde R_0}^2
R_0^{-2n} R_I^{n-1}\, ,\quad
\gamma \sim 2 \tilde R_0 R_0^{-2n} R_I^{n-1}\, .
\ee
In the model (\ref{UU2d}), the correction to the Newton law could be small
since the mass $m_\sigma$ (\ref{JGRG24}) is large and is given by
$m_\sigma^2 \sim 10^{-160 + 109 n}\,{\rm eV}^2$ in the solar system and
$m_\sigma^2 \sim 10^{-144 + 98 n}\,{\rm eV}^2$ in the air
on the earth \cite{Nojiri:2007as}.
In both cases, the mass $m_\sigma$ is very large if $n\geq 2$.

As a model corresponding to (\ref{Uf1}), we proposed
the theory \cite{Cognola:2007zu}
\be
\label{tan7}
f(R) = -\alpha_0 \left( \tanh \left(\frac{b_0\left(R-R_0\right)}{2}\right)
+ \tanh \left(\frac{b_0 R_0}{2}\right)\right) \nn
 -\alpha_I \left( \tanh \left(\frac{b_I\left(R-R_I\right)}{2}\right)
+ \tanh \left(\frac{b_I R_I}{2}\right)\right)\, .
\ee
One now assumes
\be
\label{tan8}
R_I\gg R_0\, ,\quad \alpha_I \gg \alpha_0\, ,\quad b_I \ll b_0\, ,
\ee
and
\be
\label{tan8b}
b_I R_I \gg 1\, .
\ee
When $R\to 0$ or $R\ll R_0,\, R_I$, $f(R)$ behaves as
\be
\label{tan9}
f(R) \to - \left(\frac{\alpha_0 b_0 }{2\cosh^2 \left(\frac{b_0 R_0}{2}\right) }
+ \frac{\alpha_I b_I }{2\cosh^2 \left(\frac{b_I R_I}{2}\right) }\right)R\, .
\ee
and $f(0)=0$ again. When $R\gg R_I$, it follows
\be
\label{tan10}
f(R) \to - 2\Lambda_I \equiv
 -\alpha_0 \left( 1 + \tanh \left(\frac{b_0 R_0}{2}\right)\right)
 -\alpha_I \left( 1 + \tanh \left(\frac{b_I R_I}{2}\right)\right)
\sim -\alpha_I \left( 1 + \tanh \left(\frac{b_I R_I}{2}\right)\right)\, .
\ee
On the other hand, when $R_0\ll R \ll R_I$, one gets
\be
\label{tan11}
f(R) \to -\alpha_0 \left[ 1 + \tanh \left(\frac{b_0 R_0}{2}\right)\right]
 - \frac{\alpha_I b_I R}{2\cosh^2 \left(\frac{b_I R_I}{2}\right) }
\sim -2\Lambda_0 \equiv -\alpha_0 \left[ 1 + \tanh \left(\frac{b_0
R_0}{2}\right)\right] \, .
\ee
Here, we have assumed the condition (\ref{tan8b}). One also finds
\be
\label{tan12}
f'(R)= - \frac{\alpha_0 b_0 }{2\cosh^2 \left(\frac{b_0 \left(R -
R_0\right)}{2}\right) }
- \frac{\alpha_I b_I }{2\cosh^2 \left(\frac{b_I \left(R - R_I\right)}{2}\right)
}\, ,
\ee
which has two valleys when $R\sim R_0$ or $R\sim R_I$. When $R= R_0$, \be
\label{tan13}
f'(R_0)= - \alpha_0 b_0 - \frac{\alpha_I b_I }{2\cosh^2 \left(\frac{b_I
\left(R_0 - R_I\right)}{2}\right) }
> - \alpha_I b_I - \alpha_0 b_0 \, .
\ee
On the other hand, when $R=R_I$, it follows
\be
\label{tan14}
f'(R_I)= - \alpha_I b_I - \frac{\alpha_0 b_0 }{2\cosh^2
\left(\frac{b_0 \left(R_0 - R_I\right)}{2}\right) }
> - \alpha_I b_I - \alpha_0 b_0 \, .
\ee
Due to the condition (\ref{FR1}) to avoid the anti-gravity period, one obtains
\be
\label{tan15}
\alpha_I b_I + \alpha_0 b_0 < 2\, .
\ee
In the solar system domain, on or inside the earth, where $R\gg R_0$,
$f(R)$ can be approximated by
\be
\label{tan16}
f(R) \sim -2 \Lambda_{\rm eff} + 2\alpha \e^{-b(R-R_0)}\, .
\ee
On the other hand, since $R_0\ll R \ll R_I$, by assuming Eq.~(\ref{tan8b}),
$f(R)$ (\ref{tan7}) could be also approximated by
\be
\label{tan17}
f(R) \sim -2 \Lambda_0 + 2\alpha \e^{-b_0(R-R_0)}\, ,
\ee
which has the same expression, after having identified $\Lambda_0 =
\Lambda_{\rm eff}$ and $b_0=b$.
Hence, one may check the case (\ref{tan16}) only.
The effective mass has the following form
\be
\label{tan18}
m_\sigma^2 \sim \frac{\e^{b(R-R_0)}}{4\alpha b^2}\, ,
\ee
which could be very large, that is, $m_\sigma^2 \sim 10^{1,000}\,{\rm eV}^2$
in the solar system and $m_\sigma^2 \sim 10^{10,000,000,000}\,{\rm eV}^2$ in
the air
surrounding the earth, and the correction to the Newton law becomes negligible.


One may consider another model \cite{Nojiri:2007as}:
\be
\label{VI}
f(R) = - \frac{\left(R - R_0\right)^{2k+1} + R_0^{2k+1}}{f_0 + f_1
\left\{\left(R - R_0\right)^{2k+1} + R_0^{2k+1} \right\}}\, .
\ee
It has been shown \cite{Nojiri:2007as} that for $k\geq 10$ such modified
gravity passes
the local tests. It also unifies the early-time inflation with dark energy
epoch.
In (\ref{VI}), $R_0$ is current curvature $R_0\sim
\left(10^{-33}\,\mathrm{eV}\right)^2$.
We also require
\be
\label{Uf7}
f_0 \sim \frac{R_0^{2n}}{2} \, ,\quad
f_1=\frac{1}{\Lambda_i}\, .
\ee
Here $\Lambda_i$ is the effective cosmological constant in the inflation epoch.
When $R\gg \Lambda_i$, $f(R)$ (\ref{VI}) behaves as
\be
\label{VIII}
f(R) \sim - \frac{1}{f_1} + \frac{f_0}{f_1^2 R^{2n+1}}\, .
\ee
The trace equation, which is the trace part of (\ref{JGRG13}) looks as:
\be
\label{Scalaron}
3\Box f'(R)= R+2f(R)-Rf'(R)-\kappa^2 T_\mathrm{matter}\, .
\ee
Here $T_\mathrm{matter}$ is the trace of the matter energy-momentum tensor.
For FRW metric with flat spatial part (\ref{JGRG14}), one finds
\be
\label{R}
R \sim \left(t_0 - t\right)^{-2/\left(2n+3\right)}\, ,
\ee
which diverges at finite future time $t=t_0$.
By a similar analysis, we can show that if $f(R)$ behaves as $f(R) \sim
R^\epsilon$ for large
$R$ with a constant $\epsilon$, a future singularity appears if $\epsilon>2$ or
$\epsilon<0$, which is consistent with the analysis (\ref{HS4}) via the
scalar-tensor
form of the action.
Conversely if $2\geq \epsilon \geq 0$, the singularity does not appear.
Hence, adding the term $R^2 \tilde f(R)$, where $\lim_{R\to 0} \tilde f(R) =
c_1$,
$\lim_{R\to \infty} \tilde f(R) = c_2$, to $f(R)$ (\ref{VI}),
the future singularity (\ref{R}) disappears.

Let us observe the above situation in more detail. Now we assume
\be
\label{X}
f(R) \sim F_0 + F_1 R^\epsilon\, ,
\ee
when $R$ is large. Here $F_0$ and $F_1$ are constants where $F_0$ may
vanish but $F_1\neq 0$.
In case of (\ref{VIII})
\be
\label{XI}
F_0 = - \frac{1}{f_1} \, ,\quad
F_1 = \frac{f_0}{f_1^2} \, ,\quad
\epsilon = -\left(2n+1\right) \, .
\ee
Under the assumption (\ref{X}), the trace equation (\ref{Scalaron}) gives
\be
\label{XII}
3 F_1 \Box R^{\epsilon -1} = \left\{
\begin{array}{ll} R & \ \mbox{when $\epsilon<0$ or $\epsilon=2$} \\
\left(2-\epsilon\right) F_1 R^\epsilon & \ \mbox{when $\epsilon>1$ or
$\epsilon\neq 2$}
\end{array} \right. \, .
\ee
In the FRW background with flat spatial part (\ref{JGRG14}),
when the Hubble rate has a singularity as
\be
\label{XIII}
H \sim \frac{h_0}{\left(t_0 - t\right)^\beta}\, ,
\ee
with constants $h_0$ and $\beta$, the scalar curvature $R=6\dot H + 12 H^2$
behaves as
\be
\label{XIV}
R \sim \left\{ \begin{array}{ll}
\frac{12h_0^2}{\left(t_0 - t\right)^{2\beta}} & \ \mbox{when $\beta>1$} \\
\frac{6 h_0 + 12 h_0^2}{\left(t_0 - t\right)^2} & \ \mbox{when $\beta=1$} \\
\frac{6\beta h_0}{\left(t_0 - t\right)^{\beta + 1}} & \ \mbox{when $\beta<1$}
\end{array} \right. \, .
\ee
In (\ref{XIII}) or (\ref{XIV}), $\beta\geq 1$ case corresponds to
Type I (Big Rip) singularity in \cite{Nojiri:2005sx},
$1>\beta>0$ to Type III, $0>\beta>-1$ to Type II, and $\beta<-1$
but $\beta\neq\mbox{integer}$ to Type IV.

The classification of finite-time future singularities used above is
given in ref.~\cite{Nojiri:2005sx}:
\begin{itemize}
\item Type I (``Big Rip'') : For $t \to t_s$, $a \to \infty$,
$\rho_\mathrm{eff} \to \infty$ and $\left|p_\mathrm{eff}\right| \to \infty$.
This also includes the case of $\rho_\mathrm{eff}$, $p_\mathrm{eff}$ being
finite at $t_s$.
\item Type II (``sudden'') : For $t \to t_s$, $a \to a_s$,
$\rho_\mathrm{eff} \to \rho_s$ and $\left|p_\mathrm{eff}\right| \to \infty$
\item Type III : For $t \to t_s$, $a \to a_s$,
$\rho_\mathrm{eff} \to \infty$ and $\left|p_\mathrm{eff}\right| \to \infty$
\item Type IV : For $t \to t_s$, $a \to a_s$,
$\rho_\mathrm{eff} \to 0$, $\left|p_\mathrm{eff}\right| \to 0$ and higher
derivatives of $H$ diverge.
This also includes the case in which $p_\mathrm{eff}$ ($\rho_\mathrm{eff}$)
or both of $p_\mathrm{eff}$ and $\rho_\mathrm{eff}$
tend to some finite values, while higher derivatives of $H$ diverge.
\end{itemize}
Here $\rho_\mathrm{eff} $ and $p_\mathrm{eff}$ are defined by
\be
\label{IV}
\rho_\mathrm{eff} \equiv \frac{3}{\kappa^2} H^2 \, , \quad
p_\mathrm{eff} \equiv - \frac{1}{\kappa^2} \left( 2\dot H + 3 H^2 \right)\, .
\ee

By substituting (\ref{XIV}) into (\ref{XII}), one finds that there are two
classes of consistent solutions.
The first solution is specified by $\beta =1$ and $\epsilon>1$ but
$\epsilon\neq 2$ case, which corresponds to the Big Rip
($h_0>0$ and $t<t_0$) or Big Bang ($h_0<0$ and $t>t_0$) singularity at $t=t_0$.
Another one is $\epsilon<1$, and
$\beta = - \epsilon/\left(\epsilon - 2\right)$ ($-1<\beta<1$) case, which
corresponds to (\ref{R}) and to the II Type future singularity.
In fact, we find $\epsilon = - 2n -1$ and therefore $\beta + 1 = - 2/(2n+3)$.
We should note that when $\epsilon=2$, that is, $f(R) \sim R^2$,
there is no any singular solution.
Therefore if we add the above
term $R^2 \tilde f(R)$, where $\lim_{R\to 0} \tilde f(R) = c_1$,
$\lim_{R\to \infty} \tilde f(R) = c_2$, to $f(R)$ in (\ref{VI}),
the added term dominates and
modified $f(R)$ behaves as $f(R)\sim R^2$,
the future singularity (\ref{R}) disappears.
We also note that if we add $R^n$-term with $n=3,4,5,\cdots$, the singularity
becomes (in some sense)
worse since this case corresponds to $\epsilon=n>1$, that is Big Rip case.
Using the potential which appears when we transform $F(R)$-gravity to
scalar-tensor theory \cite{Nojiri:2008fk}, it has been found that
the future singularity
may not appear in case $0<\epsilon<2$.

Thus, in order to avoid the finite-tine future singularity one has to add
$R^2$-term to above viable unification models:
\bea
\label{UU2dR2}
f(R) &=& \frac{\alpha R^{2n} - \beta R^n}{1 + \gamma R^n} + c R^2\, ,\\
\label{tan7R2}
f(R) &=& -\alpha_0 \left( \tanh \left(\frac{b_0\left(R-R_0\right)}{2}\right)
+ \tanh \left(\frac{b_0 R_0}{2}\right)\right) 
 -\alpha_I \left( \tanh \left(\frac{b_I\left(R-R_I\right)}{2}\right)
+ \tanh \left(\frac{b_I R_I}{2}\right)\right) + c R^2\, , \\
\label{VIR2}
f(R) &=& - \frac{\left(R - R_0\right)^{2k+1} + R_0^{2k+1}}{f_0 + f_1
\left\{\left(R - R_0\right)^{2k+1} + R_0^{2k+1} \right\}} + c R^2\, .
\eea
The addition of this term has
been proposed first in ref.~\cite{Abdalla:2004sw} where it was shown that in
this case
the Big Rip singularity disappears.
Moreover, such term which effectively cures future singularity also supports
the early-time inflation.
In other words, adding such $R^2$-term to gravitational dark energy model may
lead to emergence of inflationary phase in the model as it was first observed
in ref.~\cite{Nojiri:2003ft}. In case, when model already contains the
inflationary era, its dynamics will be changed by $R^2$-term.
The investigation which shows that it cures all types of future singularity has
been done in refs.~\cite{Nojiri:2008fk,bamba}.
In fact, it was realized lately that some phenomenological
problems \cite{others} of $F(R)$-dark energy (like consistent description of
neutron stars) may be resolved in the presence of $R^2$-term.
Moreover, the traditional phantom/quintessence (fluid/scalar) dark energy
models often bring the
universe to finite-time future singularity.
It has been demonstrated in ref.~\cite{Nojiri:2009pf}
   that the natural prescription to cure singularity in this models is again
the
addition of $R^2$-term.
In other words, curing future singularity requests from specific dark energy
model to be (at least, partly) the modified gravity!


\subsection{New viable $F(R)$-models}

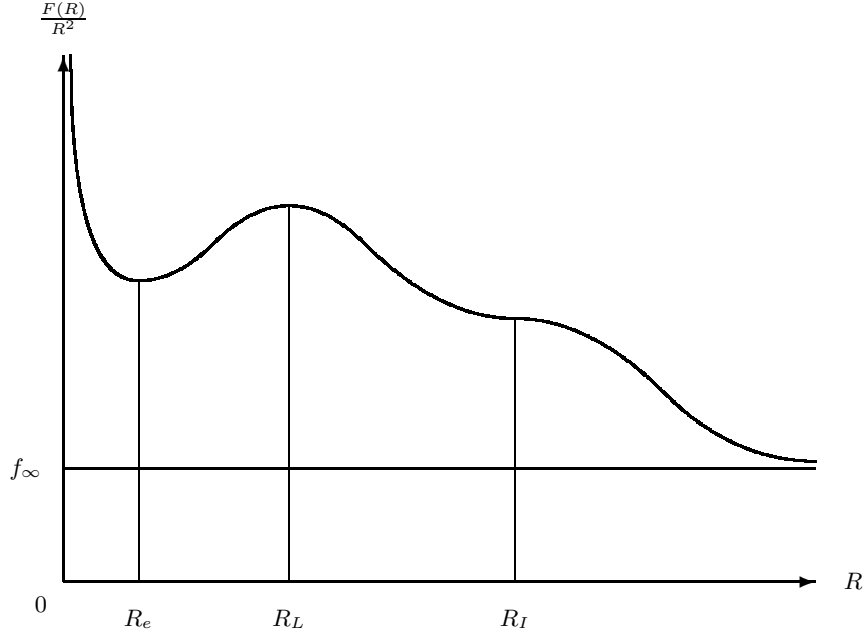
\begin{figure}

\begin{center}

\unitlength=1mm

\begin{picture}(120,100)

\put(7,7){\makebox(0,0){$0$}}
\put(10,85){\makebox(0,0){$\frac{F(R)}{R^2}$}}
\put(115,10){\makebox(0,0){$R$}}
\put(5,25){\makebox(0,0){$f_\infty$}}
\put(20,5){\makebox(0,0){$R_e$}}
\put(40,5){\makebox(0,0){$R_L$}}
\put(70,5){\makebox(0,0){$R_I$}}

\thicklines

\put(10,10){\vector(1,0){100}}
\put(10,10){\vector(0,1){70}}

\qbezier(11,80)(11,50)(20,50)
\qbezier(20,50)(25,50)(30,55)
\qbezier(30,55)(35,60)(40,60)
\qbezier(40,60)(45,60)(50,55)
\qbezier(50,55)(60,45)(70,45)
\qbezier(70,45)(80,45)(90,35)
\qbezier(90,35)(99,26)(110,26)

\thinlines

\put(10,25){\line(1,0){100}}
\put(20,10){\line(0,1){40}}
\put(40,10){\line(0,1){50}}
\put(70,10){\line(0,1){35}}

\end{picture}

\end{center}

\caption{\label{fig1} The qualitative behavior of $\frac{F(R)}{R^2}$
versus $R$ for a viable model.}
\end{figure}

The above analysis shows that, in order to obtain a realistic and viable model,
$F(R)$-gravity should satisfy the following conditions:
\begin{enumerate}
\item\label{req1} When $R\to 0$,
the Einstein gravity is recovered, that is,
\be
\label{E1}
F(R) \to R \quad \mbox{that is,} \quad \frac{F(R)}{R^2} \to \frac{1}{R}\, .
\ee
This also means that there is a flat space solution as in (\ref{Uf4}).
\item\label{req2} As discussed after Eq.~(\ref{dS6}), there appears stable de
Sitter solution,
which corresponds to the late-time acceleration and therefore the curvature is
small
$R\sim R_L \sim \left( 10^{-33}\, \mathrm{eV}\right)^2$.
This requires, when $R\sim R_L$,
\be
\label{E2}
\frac{F(R)}{R^2} = f_{0L} - f_{1L} \left( R - R_L \right)^{2n+2}
+ o \left( \left( R - R_L \right)^{2n+2} \right)\, .
\ee
Here $f_{0L}$ and $f_{1L}$ are positive constants and $n$ is a positive
integer.
\item\label{req3} As also discussed after Eq.~(\ref{dS6}), there appears
quasi-stable
de Sitter solution, which corresponds to the inflation
in the early universe and therefore the curvature is large
$R\sim R_I \sim \left( 10^{16 \sim 19}\, \mathrm{GeV}\right)^2$. The de Sitter
space should
not be exactly stable so that the curvature decreases very slowly. This require
\be
\label{E3}
\frac{F(R)}{R^2} = f_{0I} - f_{1I} \left( R - R_I \right)^{2m+1}
+ o \left( \left( R - R_I \right)^{2m+1} \right)\, .
\ee
Here $f_{0I}$ and $f_{1I}$ are positive constants and $m$ is a positive
integer.
\item\label{req4} Following the discussion after (\ref{HS3}), when $R\to
\infty$,
in order to avoid the curvature singularity, it is proposed
\be
\label{E4}
F(R) \to f_\infty R^2 \quad \mbox{that is} \quad \frac{F(R)}{R^2} \to f_\infty
\, .
\ee
Here $f_\infty$ is a positive and sufficiently small constant.
Instead of (\ref{E4}), we may take
\be
\label{E5}
F(R) \to f_{\tilde \infty} R^{2 - \epsilon} \quad \mbox{that is}
\quad \frac{F(R)}{R^2} \to \frac{f_{\tilde\infty}}{R^\epsilon} \, .
\ee
Here $f_{\tilde\infty}$ is a positive constant and $0< \epsilon <1$.
The above condition (\ref{E4}) or (\ref{E5}) prevents both of the future
singularity
and the singularity at high density matter.
\item\label{req5} As in (\ref{FR1}), in order to avoid the anti-gravity, we
require
\be
\label{E6}
F'(R)>0\, ,
\ee
which is rewritten as
\be
\label{E7}
\frac{d}{dR} \left( \ln \left( \frac{F(R)}{R^2} \right)\right)
> - \frac{2}{R}\, .
\ee
\item\label{req6}
Combining the conditions (\ref{E1}) and (\ref{E6}), one finds
\be
\label{E8}
F(R)>0\, .
\ee
\end{enumerate}
The conditions \ref{req1} and \ref{req2} show that there must appear unstable
extra de Sitter
solution at $R=R_e$ $\left( 0< R_e < R_L \right)$, see Fig.\ref{fig1}.
Since the  universe evolution will stop at $R=R_L$ since the de Sitter
solution $R=R_L$ is
stable, the curvature never becomes smaller than $R_L$ and therefore the extra
de Sitter solution
is not realized.
The behavior of $\frac{F(R)}{R^2}$ which satisfies the conditions (\ref{E1}),
(\ref{E2}),
(\ref{E3}), (\ref{E4}), and (\ref{E8}) is given in Fig.\ref{fig1}.

An example of such $F(R)$-gravity is
\bea
\label{EE1}
\frac{F(R)}{R^2} &=& \left\{ \left(X_m \left(R_I;R\right) -
X_m\left(R_I;R_1\right) \right)
\left(X_m\left(R_I;R\right) - X_m\left(R_I;R_L\right) \right)^{2n+2} \right.
\nn
&& \left. + X_m\left(R_I;R_1\right) X_m\left(R_I;R_L\right) ^{2n+2}
+ f_\infty^{2n+3} \right\}^{\frac{1}{2n+3}} \, , \nn
X_m\left(R_I;R\right) &\equiv & \frac{\left(2m +1\right) R_I^{2m}}{\left( R -
R_I \right)^{2m+1} + R_I^{2m+1}} \, .
\eea
Here $n$ and $m$ are integers greater or equal to unity, $n,m\geq 1$ and
$R_1$ is a parameter related with $R_e$ by
\be
\label{EE2}
X\left(R_I;R_e\right) = \frac{\left(2n+2\right)
X\left(R_I;R_1\right)X\left(R_I;R_1\right)
+ X\left(R_I;R_L\right)}{2n+3}\, ,
\ee
It is assumed
\be
\label{EEE1}
0<R_1<R_L \ll R_I \, .
\ee
  $X\left(R_I;R_e\right)$ is a monotonically decreasing function of $R$
and in the limit $R\to 0$, $X\left(R_I;R_e\right)$ behaves as
\be
\label{EE3}
X\left(R_I;R_e\right) \to \frac{1}{R}\, ,
\ee
which tells that in the limit, $F(R)$ (\ref{EE1}) is (\ref{E1}) and therefore
the condition \ref{req1} is satisfied.
On the other hand, when $R\to \infty$,
\be
\label{EE4}
X\left(R_I;R_e\right) \to \frac{\left(2m+1\right) R_I^{2m}}{R^{2m+1}} \to 0\, ,
\ee
and therefore $F(R)$ behaves as (\ref{E4}) and the condition \ref{req4}
is satisfied.

When $R\sim R_L$, we find the behavior of (\ref{E2}), where
\bea
\label{EE5}
f_{0L} &=& \left\{ X_m\left(R_I;R_1\right) X_m\left(R_I;R_L\right)^{2n+2}
+ f_\infty^{2n+3} \right\}^{\frac{1}{2n+3}} \, , \nn
f_{1L} &=& \frac{1}{2m+3} \left\{ X_m\left(R_I;R_1\right)
X_m\left(R_I;R_L\right)^{2n+2}
+ f_\infty^{2n+3} \right\}^{- \frac{2\left(n+1\right)}{2n+3}}
\left(X_m \left(R_I;R_1\right) - X_m\left(R_I;R_L\right) \right) \nn
&& \times \frac{\left( 2m + 1 \right)^{4\left(m+1\right)}
\left\{ R_I \left(R_L - R_I\right)\right\}^{4m\left(m+1\right)} }
{\left\{ \left( R_L - R_I \right)^{2m+1} + R_I^{2m+1}
\right\}^{4\left(m+1\right)}} \, .
\eea
Then the condition \ref{req2} is satisfied.

On the other hand, when $R\sim R_I$, we also find the behavior of (\ref{E3}),
where
\bea
\label{EE6}
f_{0I} &=& \left\{ \left(X_m \left(R_I;R_I\right) - X_m\left(R_I;R_1\right)
\right)
\left(X_m\left(R_I;R_I\right) - X_m\left(R_I;R_L\right) \right)^{2n+2} \right.
\nn
&& \left. + X_m\left(R_I;R_1\right) X_m\left(R_I;R_L\right) ^{2n+2}
+ f_\infty^{2n+3} \right\}^{\frac{1}{2n+3}} \nn
f_{1I} &=& \frac{2m+1}{R_I^{2m+2}}
\left\{ \left(X_m \left(R_I;R_I\right) - X_m\left(R_I;R_1\right) \right)
\left(X_m\left(R_I;R_I\right) - X_m\left(R_I;R_L\right) \right)^{2n+2} \right.
\nn
&& \left. + X_m\left(R_I;R_1\right) X_m\left(R_I;R_L\right) ^{2n+2}
+ f_\infty^{2n+3} \right\}^{\frac{2\left(n+1\right)}{2n+3}} \left\{
\left(X_m\left(R_I;R_I\right) - X_m\left(R_I;R_L\right) \right)^{2n+2} \right.
\nn
&& \left. + \left(2n+2\right)\left(X_m \left(R_I;R_I\right) -
X_m\left(R_I;R_1\right) \right)
\left(X_m\left(R_I;R_I\right) - X_m\left(R_I;R_L\right) \right)^{2n+1}
\right\}\, .
\eea
Hence, the condition \ref{req3} is satisfied.

We now investigate the mass of the scalar field $\sigma$ in order to check
if the Chameleon mechanism \cite{khoury} works or not. For this purpose, one
investigates the region
$R_1<R_L \ll R \ll R_I$. In this region, $X_m\left(R_I;R\right)$ can be
approximated as
\be
\label{EE7}
X_m\left(R_I;R\right) \sim \frac{1}{R}\, ,\quad
X_m\left(R_I;R_1\right) \sim \frac{1}{R_1}\, ,\quad
X_m\left(R_I;R_L\right) \sim \frac{1}{R_L}\, .
\ee
Then $F(R)$ can be approximated as
\be
\label{EE8}
\frac{F(R)}{R^2} \sim f_\infty + \frac{f_n}{R}\, ,\quad
f_n \equiv \frac{R_1 + \left(2n+2\right) R_L f_\infty^{-2n -2}}
{\left(2n+3\right) R_1 R_L^{2n+2}}
\sim \frac{f_\infty^{-2n -2}}{R_L^{2n+2}}\, .
\ee
In the last equation, it was assumed $R_1\sim R_L$.
If we assume $f_\infty R \ll f_n$ or
\be
\label{EE9}
\frac{1}{f_\infty^{2n+3}} \gg R R_L^{2n+2}\, ,
\ee
one gets
\be
\label{EE10}
m_\sigma^2 \sim \frac{3}{4f_\infty}\, .
\ee
If we consider the region inside the earth, since $1\,{g}\sim 6\times
10^{32}\,\mathrm{eV}$
and $1\,\mathrm{cm}\sim \left(2\times 10^{-5}\,\mathrm{eV}\right)^{-1}$,
the density is about $\rho\sim 1 \mathrm{g/cm^3} \sim 5\times 10^{18}\,
\mathrm{eV}^4$.
This shows that the magnitude of the curvature is
$R \sim \kappa^2 \rho \sim \left(10^{-19}\,\mathrm{eV}\right)^2$.
In the air on the earth, one finds $\rho\sim 10^{-6} \mathrm{g/cm^3} \sim
10^{12}\,
\mathrm{eV}^4$,
which gives $R_0 \sim \kappa^2 \rho \sim \left(10^{-25}\,\mathrm{eV}\right)^2$.
In the solar system, there could be interstellar gas. Typically, in the
interstellar gas, there is one proton (or hydrogen atom) per
$1\,\mathrm{cm}^3$,
which shows $\rho\sim 10^{-5}\,\mathrm{eV}^4$, $R_0\sim 10^{-61}\,
\mathrm{eV}^2$.
Then, the condition (\ref{EE9}) can be easily satisfied, for example, we can
chose $\frac{1}{f\infty}\sim \mathrm{MeV}^2$. Then the Compton length
of the scalar field becomes very small and the correction to the Newton law is
negligible.

There remains the condition \ref{req5}.
Since $F(R)$ (\ref{EE1}) satisfies Eqs. (\ref{E1}) and (\ref{E4}), it is clear
that
Eq.~(\ref{E6}) or (\ref{E7}) and therefore the condition \ref{req5} are
satisfied when $R\to 0$ or
$R\to \infty$. Since $\frac{F(R)}{R^2}$ is a monotonically increasing function
of $R$ in the region
$R_e<R<R_L$, Eq.~(\ref{E6}) or (\ref{E7}) is trivially satisfied in the region.
In the region $R_1<R_L \ll R \ll R_I$, since $\frac{F(R)}{R^2}$ behaves as
(\ref{EE8}), we find
Eq.~(\ref{E6}) or (\ref{E7}) is satisfied again.
Then the condition \ref{req5} seems to be satisfied in all the region.

After the inflation at $R=R_I$, the radiation and the matter are generated
generated.
If the energy densities of the radiation and the matter dominate compared with
the
contribution from $f(R)=R(R) - R$, that is, only the first term in (\ref{rhoE})
dominates,
the radiation/matter dominated universe will be realized.
The densities of the radiation and the matter decrease rapidly compared with
the contribution
from $f(R)$-term in the late-time and when the curvature arrives at $R\sim
R_L$, the late-time
acceleration occurs.

The condition (\ref{E5}) indicates that there appears $R^2$ term in the very
high curvature region.
The $R^2$-term will generate the inflation besides the inflation at $R=R_I$ if
the universe started
with very high curvature $R\gg R_I$. Since the inflation due to $R^2$-term
is unstable, the inflation at the very early stage would stop but when the
curvature decreases and reaches
$R\sim R_I$, there will occur the inflationary phase again.

Thus, we demonstrated that number of viable $F(R)$-gravity models may explain
the early-time inflation with dark energy epoch in unified way.

\section{Universality of viscous ratio bound in 5d $F(R)$-gravity \label{Sec7}}

In the present section, we discuss the possible applications of 5d modified
$F(R)$-gravity
as non-perturbative stringy gravity in AdS/CFT correspondence,
which has provided a gravity dual framework to analyze strongly coupled gauge
theory.
Specifically, shear viscosity in the hydrodynamical limit of the gauge theories
has been well studied \cite{Policastro:2001yc, Buchel:2003tz, Kovtun:2004de,
Dutta:2008gf}.
It was found that the ratio of the the coefficient of the shear viscosity
$\eta$ and
the entropy density $s$ should be larger than $1/4\pi$,
\be
\label{I}
\frac{\eta}{s} \geq \frac{1}{4\pi}\, ,
\ee
in the Einstein gravity. It turns out this bound is not universal.
The correction to the bound coming from the higher order corrections,
(Einstein-Gauss-Bonnet gravity or $R^3$ gravity,) has been calculated
in refs.~\cite{Brigante:2007nu,Neupane:2008dc,Cai:2009zv}.
In this section, we consider $F(R)$-gravity as a prototype non-linear model
which may come
from non-perturbative effects of string theory. We show
the universality of the viscous bound ratio emerges even in this case.
The non-universality of such bound is due to account of the effects
in the next order of loop expansion, but our results indicate that for
non-linear non-perturbative stringy model
this may be not true.

Let us start from $F(R)$-gravity in $D$ dimensions:
\be
\label{M1}
I = \int d^D x \sqrt{-g} F(R) \, .
\ee
Later one may choose $D=5$.
Here we do not consider the contribution from matter.
Then from (\ref{JGRG13}), the equation of the motion is given by
\be
\label{M2}
0 = \frac{1}{2}g_{\mu\nu} F(R) - R_{\mu\nu}F'(R) + \nabla_\mu \nabla_\nu F'(R)
   - g_{\mu\nu}\nabla^2 F'(R) \, .
\ee
If we assume the curvatures are covariantly constant
\be
\label{vs1}
R=R_0\, ,\quad R_{\mu\nu} = \frac{R_0}{D}g_{\mu\nu}\, ,
\ee
with a constant $R_0$, Eq.~(\ref{M2}) reduces to an algebraic equation:
\be
\label{vs2}
0 = \frac{D}{2} F(R_0) - R_0 F'(R_0)\, ,
\ee
which could be solved with respect to $R_0$.
This shows that any vacuum solution in the Einstein gravity like Schwarzschild
solution,
Kerr solution, gravitational wave, etc. is the solution of $F(R)$-gravity.
Especially if $R_0<0$, we find Schwarzschild-anti-de Sitter (S-AdS)
\be
\label{S-AdS}
ds_D^2= -V(r)dt^2 + V^{-1}(r)dr^2+r^2 \sum_{i=1}^{D-2}\left(dx^i\right)^2\, ,
\quad
V(r) = - \frac{M}{r^{D-3}} + \frac{r^2}{l^2}\, .
\ee
Kerr-anti-de Sitter (K-AdS),
and Black string-anti de Sitter solutions.
Here the length scale $l$ is related with $R_0$ by
\be
\label{l}
R_0 = -\frac{D(D-1)}{l^2}\, .
\ee
Eq.~(\ref{S-AdS}) shows that the horizon radius $r_0$ is given by
\be
\label{r0}
r_0 = \left( l^2 M \right)^{\frac{1}{D-1}}\, .
\ee
If the radial coordinate $r$ is redefined as
\be
\label{rho}
\rho \equiv \frac{1}{r^2}\, ,
\ee
the metric (\ref{S-AdS}) is rewritten as
\be
\label{rho1}
ds_D^2 = - \frac{1 - M l^2 \rho^{\frac{D-1}{2}}}{l^2 \rho} dt^2
+ \frac{ l^2 d\rho^2}{4 \rho^2 \left( 1 - M l^2 \rho^{\frac{D-1}{2}} \right)}
+ \frac{1}{\rho} \Omega^2\, .
\ee

We now expand the action (\ref{M1}) from the solution
$g_{\mu\nu} = g^{(0)}_{\mu\nu}$ (\ref{vs2})
\be
\label{vs3}
g_{\mu\nu} = g^{(0)}_{\mu\nu} + h_{\mu\nu}\, .
\ee
Up to the third order in $h_{\mu\nu}$:
\bea
\label{vs4}
S &=& \int d^D x \sqrt{-g} \left[ F(R_0) - \frac{1}{4} F(R_0) h^{\mu\nu}
h_{\mu\nu}
 - \frac{1}{2} \left( - \frac{R_0 F'(R_0)}{2D} + \frac{R_0^2 F''(R_0)}{D^2}
\right) h^2 \right. \nn
&& - \frac{1}{2} \left(F'(R_0) - \frac{R_0 F''(R_0)}{D}\right) h
\nabla^{(0)\,\mu} \nabla^{(0)\,{\mu}} h_{\mu\nu}
 - \frac{1}{2} \left(- \frac{F'(R_0)}{2} + \frac{2R_0 F''(R_0)}{D}\right) h
\left(\nabla^{(0)}\right)^2 h \nn
&& - \frac{F'(R_0)}{4} h^{\mu\nu} \left\{ \left(\nabla^{(0)}\right)^2
h_{\mu\nu}
 - \nabla^{(0)\,\rho} \left( \nabla^{(0)}_\mu h_{\nu\rho} + \nabla^{(0)}_\nu
h_{\mu\rho}\right) \right\} \nn
&& \left. - \frac{1}{2} F''(R_0)
\left\{ \nabla^{(0)\,\mu} \nabla^{(0)\,{\mu}} h_{\mu\nu} -
\left(\nabla^{(0)}\right)^2 h \right\}^2
\right] \, .
\eea
Here
\be
\label{vs5}
h \equiv {g^{(0)}}^{\mu\nu} h_{\mu\nu}\, ,\quad
h^{\mu\nu} \equiv {g^{(0)}}^{\mu\rho} {g^{(0)}}^{\nu\sigma} h_{\rho\sigma}\, ,
\ee
and $\nabla^{(0)}_\mu$ expresses the covariant derivative given by
$g^{(0)}_{\mu\nu}$.
Then the viscosity coefficient $\eta$ can be read from the coefficient in front
of
$\left(\partial_r h_{ij}\right)$ which is contained only in the 6-th term as
$- \left(F'(R_0)/4\right) h^{\mu\nu} \left(\nabla^{(0)}\right)^2 h_{\mu\nu}$
\be
\label{vs6}
\eta = F'(R_0)\, .
\ee
In case of $F(R)$-gravity, the entropy is given by \cite{Cognola:2005de}
\be
\label{vs7}
S = 4\pi A F'(R_0)\, ,
\ee
which gives the entropy density $s$ as
\be
\label{vs8}
s = \frac{S}{A} = 4\pi F'(R_0) \, .
\ee
Combining (\ref{vs8}) with (\ref{vs6}), one gets
\be
\label{vs9}
\frac{\eta}{s} = \frac{1}{4\pi}\, ,
\ee
which is not changed from the case of the Einstein gravity.
This result is completely consistent with the ones in \cite{Kats:2007mq}
and \cite{Banerjee:2009ju}.

As an example, we consider $R^2$-gravity with a cosmological constant
\be
\label{R2-1}
F(R) = \frac{1}{2\kappa^2}\left\{ R + \frac{\left( D-1 \right) \left( D - 2
\right)}{l_\Lambda^2}
+ c R^2 \right\} \, .
\ee
Here the cosmological constant $\Lambda$ is given by
\be
\label{R2-1b}
\Lambda = - \frac{\left( D-1 \right) \left( D - 2 \right)}{l_\Lambda^2}\, .
\ee
Then (\ref{vs2}) gives
\be
\label{R2-2}
R_0 = - \frac{D(D-1)}{l^2}
= - \frac{D-2}{2 c (D-4)} \left\{ 1 \pm \sqrt{ 1
   - \frac{4c D(D-1)(D-4)}{(D-2) l_\Lambda^2}}\right\}\, .
\ee
In the signs $\pm$ of (\ref{R2-2}), the minus sign $-$ reduces to the Einstein
gravity case
in $c\to 0$ limit.
When the minus sign is chosen, in the limit we find
\be
\label{R2-3}
R_0 = - \frac{D(D-1)}{l_\Lambda^2}\left\{ 1 + \frac{c D (D-1)
(D-4)}{(D-2)l_\Lambda^2}
+ \mathcal{O}\left(c^2\right) \right\}\, .
\ee
Then (\ref{vs6}) gives
\be
\label{R2-4}
\eta = \frac{1}{2\kappa^2} \left( 1 + 2c R_0\right)
\sim \frac{1}{2\kappa^2} \left( 1 - \frac{2 c D ( D-1)}{l_\Lambda^2}\right)\, .
\ee
Especially when $D=5$, one obtains
\be
\label{R2-5}
\eta \sim \frac{1}{2\kappa^2} \left( 1 - \frac{40 c }{l_\Lambda^2}\right)\, .
\ee
This result essentially agrees with the previous ones in \cite{Kats:2007mq,
Banerjee:2009ju, Dutta:2008gf}.
Note that in \cite{Banerjee:2009ju}, the horizon radius was shifted from the
Einstein gravity
case ($c=0$) by adding the correction term ($c\neq 0$), via account of the
shift
of the length parameter $l_\Lambda$ in (\ref{R2-1b}) corresponding to the
Einstein gravity to
$l$ in (\ref{R2-2}), which includes the correction coming from $c$.
The shift can be found by using (\ref{R2-2}) and (\ref{R2-3}).
\be
\label{R2-6}
l = l_\Lambda\left\{ 1 - \frac{c D (D-1) (D-4)}{2(D-2)l_\Lambda^2}
+ \mathcal{O}\left(c^2\right) \right\}\, ,
\ee
which gives, in $D=5$,
\be
\label{R2-7}
l = l_\Lambda\left\{ 1 - \frac{10 c }{3 l_\Lambda^2}
+ \mathcal{O}\left(c^2\right) \right\}\, .
\ee
When $D=5$, Eq.~(\ref{r0}) shows that $r_0$ is given by $r_0 = l^{\frac{1}{2}}
M^{\frac{1}{4}}$.
Then the horizon radius $r_0$ may be shifted as $r_0 \to
\left(l/l_\Lambda\right)^{\frac{1}{2}} r_0$
by the correction from $cR^2$ term and therefore the
area $A_0$ of the horizon as $A_0\to \left(l/l_\Lambda\right)^{\frac{3}{2}}
A_0$.
Since the quantity $\eta$ is the density on
the horizon, we may regard $\eta$ is also shifted as
\be
\label{R2-8}
\eta \to \left(l/l_\Lambda\right)^{\frac{3}{2}} \eta
= \frac{1}{2\kappa^2} \left( 1 - \frac{45 c }{l_\Lambda^2}\right)\, ,
\ee
which completely agrees with the previous ones
in refs.~\cite{Kats:2007mq, Banerjee:2009ju, Dutta:2008gf}.
We should note, however, that since $M$ in (\ref{S-AdS}) is the constant of the
integration
the shift of the horizon radius can be always absorbed into the redefinition of
$M$.
Therefore in this section, we may fix the value of the radius, which makes
difference
in the expression of the viscosity coefficient
$\eta$ but as we have seen, the ratio of $\eta$ and the entropy density in
(\ref{vs9}) does not
depend on the radius and exactly agrees
with the results in \cite{Kats:2007mq, Banerjee:2009ju, Dutta:2008gf}.
This is because both of the viscosity coefficient $\eta$ and the entropy
density $s$ are
densities, they shift in an identical way by the shift of the horizon. For
example, if we change the constant
of the integration from $M_1$ to $M_2$, $\eta$ and $s$ shift as
\be
\label{R2-9}
\eta \to \left(M_2/M_1\right)^{\frac{3}{4}} \eta\, ,\quad
s \to \left(M_2/M_1\right)^{\frac{3}{4}} s \, .
\ee
Then in the viscous ratio the shift is canceled with each other.
Hence, we proved the emergence of the universality viscous ratio bound for 5d
$F(R)$-gravity.
This may be understood also due to the fact of mathematical equivalence of
$F(R)$-gravity
with scalar-tensor theory which also satisfies to universal viscous ratio
bound.
Nevertheless, other non-perturbative stringy gravity models (like $F(G)$
gravity \cite{Nojiri:2005jg}) should
be investigated in order to understand if the restoration of universality
of viscous ratio bound is the common property of non-perturbative models.

\section{Discussion \label{Sec8}}

In summary, brief introduction to cosmological aspects of $F(R)$-gravity is 
made. Spatially-flat FRW equations and its typical accelerating solutions are 
discussed. The generalized (gravitational) fluid as well as scalar-tensor 
description of the theory under investigation is given. 
The emergence of the Newton law in scalar-tensor formulation is described. 
Cosmological reconstruction technique for $F(R)$-gravity is overviewed as well 
as general description of stability condition for cosmological solution. The 
review of known viable $F(R)$-gravities unifying the inflation with dark energy 
epoch is given. Some of such models may lead to finite-time future 
singularities which may be cured by the addition of $R^2$-term relevant at high 
curvature. It is remarkable that this term not only cures future singularity 
but also contributes to realization of inflationary era. In fact, the 
alternative gravity dark energy model which does not describe inflation may 
also lead to inflationary phase after addition of $R^2$-term. 

New realistic non-singular $F(R)$-gravity unifying inflation with dark energy 
is introduced and studied in detail. It is shown that inflationary epoch is 
unstable and decays due to gravitational action structure.
Hence, modified gravity provides not only the universal unification scenario 
but also the gravitational exit from inflation.

Section \ref{Sec6} addresses different, five-dimensional application of $F(R)$-gravity 
as non-perturbative stringy gravity within AdS/CFT correspondence. It is less 
related with previous sections which are devoted to cosmological study of the 
four-dimensional theory. It is demonstrated here that
the well-known viscous ratio bound in five-dimensional $F(R)$-gravity is 
universal unlike to the Einstein-Gauss-Bonnet gravity or stringy gravity with 
higher-order perturbative gravitational corrections.

Modified gravity unifying inflation with late-time acceleration has passed 
number of different checks, related with theoretical considerations, local 
tests or observational data. It is still considered as most natural 
candidate for description of the whole universe evolution in unified way.
It is remarkable that just the same models may be used for inflation-dark 
energy unified description for totally different classes of theories. For 
instance, such unification turns out to be possible in recent 
Ho\v{r}ava-Lifshitz $F(R)$-gravity where unified models were introduced in 
ref.~\cite{masud}.

\section*{Acknowledgments \label{Ack}}

We are indebted to R. Myers and M. Takeuchi for the discussion.
We are grateful to M. Sasaki for kind hospitality during YKIS 
International Conference Gravitation and Cosmology,YITP 2010.
This research has been supported in part
by MEC (Spain) project FIS2006-02842 and AGAUR(Catalonia) 2009SGR-994 (SDO),
by Global COE Program of Nagoya University (G07)
provided by the Ministry of Education, Culture, Sports, Science \& Technology
and by the JSPS Grant-in-Aid for Scientific Research (S) \# 22224003 (SN).

\end{document}